\documentclass[12pt,journal, onecolumn, draftcls]{IEEEtran}
\usepackage{amsmath}
\usepackage{graphicx}
\usepackage{psfrag}
\usepackage{array}
\usepackage{amssymb}
\usepackage{color}

\usepackage{amsmath,graphicx}
\usepackage{amssymb}
\usepackage{hyperref}
\usepackage{color}
\usepackage{subfigure}
\usepackage{algorithmic}
\usepackage{algorithm}
\usepackage{psfrag}
\usepackage{amsthm}
\usepackage{flushend}

\begin{document}

\title{Joint Robust Transmit/Receive Adaptive Beamforming for MIMO Radar Using Probability-Constrained Optimization}
\author{Weiyu Zhang, and Sergiy~A.~Vorobyov, 
{\it Senior Member, IEEE}

\thanks{
W.~Zhang is with the State Key Laboratory of Acoustics,
Institute of Acoustics, Chinese Academy of Sciences, Beijing 100190, China.
His work is supported by China Scholarship Council. He is also with the Department of Signal Processing and Acoustics, Aalto University, Espoo, FI-00076, Finland  (email: zwy@mail.ioa.ac.cn).

S.~A.~Vorobyov is with the Department of Signal Processing and Acoustics, Aalto University, Espoo, FI-00076, Finland  (email: svor@ieee.org).
}}

\maketitle

\begin{abstract}
In this letter, a joint robust transmit/receive adaptive beamforming for multiple-input multiple-output (MIMO) radar based on probability-constrained optimization approach is developed in the case of Gaussian and arbitrary distributed mismatch present in both the transmit and receive signal steering vectors. A tight lower bound of the probability constraint is also derived by using duality theory. The formulated probability-constrained robust beamforming problem is nonconvex and NP-hard. However, we reformulate its cost function into a bi-quadratic function while the probability constraint splits into transmit and receive parts. Then, a block coordinate descent method based on second-order cone programming is developed to address the biconvex problem. Simulation results show an improved robustness of the proposed beamforming method as compared to the worst-case and other existing state-of-the-art joint transmit/receive robust adaptive beamforming methods for MIMO radar.
\end{abstract}


\section{Introduction}
Multiple-input multiple-output (MIMO) radar has become the focus of intensive research \cite{MIMO_radar_book}-\cite{Hassanien5}. MIMO radar with colocated antennas allows to form a virtual array with a larger number of virtual antenna elements. It significantly enhances resolution \cite{Bliss} and parameter identifiability \cite{MIMO diversity}, \cite{on parameter}, allows for direct applicability of adaptive arrays for target detection  \cite{MIMO diversity}, \cite{Xu1}, \cite{Bekkerman}, and enhances flexibility for transmit beampattern design \cite{Fuhrmann}-\cite{Browning}.

Robust adaptive beamforming has been widely used to achieve higher resolution capability in the traditional frame work of phased receive array \cite{VanTrees}-\cite{Vorobyov1}. The well-known sample matrix inversion (SMI) beamformer \cite{VanTrees}-\cite{Gershman} has been recently used in the MIMO radar (joint transmit/receive) beamforming context in \cite{Hassanien3}, \cite{Hassanien4}. However, the SMI beamformer does not provide sufficient robustness against a mismatch between the presumed and actual transmit and receive steering vectors. The so-called loaded SMI (LSMI) \cite{Besson1}, \cite{Carlson}, \cite{diagonal loading} and the worst-case optimization-based beamformers \cite{Vorobyov}-\cite{Ripple} can be used, but they may be overly conservative in practical applications, because the actual worst operational conditions may occur with a rather low
probability. Robust adaptive beamforming approaches developed in \cite{Vorobyov3}-\cite{Vorobyov6} for traditional phased arrays are then more appropriate since they are designed to guarantee the robustness against the signal steering vector mismatch with a certain selected probability.

In this letter, we propose a joint robust transmit/receive adaptive beamforming for MIMO radar using the probability-constrained optimization  and provide a mathematical analysis of the tight lower bound for the probability constraint by using duality theory for the case of arbitrary distributed mismatch. We transfer the primal nonconvex optimization problem into a biconvex problem \cite{Gorski} and address it by using block coordinate descent (BCD) approach \cite{Tseng}-\cite{Zou}.

\section{Signal Model}
Consider a MIMO radar system equipped with $M_{\rm t}$ transmit and $M_{\rm r}$ receive antenna elements. Both the transmit and receive arrays are assumed to be closely located so that they share the same spatial angle of a far-field target. Let ${\boldsymbol \phi}(t) \triangleq [\phi_1(t),\ldots,\phi_{M_{\rm t}}(t)]^T$ be the waveform vector that contains the complex envelopes of orthogonal waveforms which are emitted by different transmit antennas, i.e., $\int_T\phi_i(t)\phi_j^\ast(t)dt=\delta(i-j), i,j=1,\ldots,M_{\rm t},$ where $T$ is the pulsewidth and $\delta(\cdot)$ is the Kronecker delta function. Here $(\cdot)^T$ and $(\cdot)^\ast$ stand for the transpose and complex conjugate operations, respectively.

Assuming that $k$ targets are present, the $M_{\rm r}\times1$ received complex observation vector can be written as
\begin{equation}
\mathbf{x}(t,\tau)=\displaystyle\sum_{i=1}^k\beta_i(\tau) \left[  \mathbf{a}_{\rm t}^T(\theta_i){\boldsymbol \phi}(t) \right]  \mathbf{a}_{\rm r}(\theta_i)+\mathbf{n}(t,\tau)
\end{equation}
where $\tau$ is the slow time index, i.e., the pulse number, $\beta_i(\tau)$ is the reflection coefficient of the $i$th source with variance $\sigma_\beta^2$, $\mathbf{a}_{\rm t}(\theta)$ and $\mathbf{a}_{\rm r}(\theta)$ are the transmit and receive steering vectors, respectively, and $\mathbf{n}(t,\tau)$ is zero-mean white Gaussian noise. The reflection coefficients $\beta_i(\tau), i=1,\ldots,k$ are assumed to remain constant during the whole pulse, but vary independently from pulse to pulse.

By matched filtering the received data to the $M_{\rm t}$ orthogonal waveforms at the receiving end and stacking the individual vector components in one column vector, the  $M_{\rm t}M_{\rm r}\times1$ virtual data vector can be obtained as
\begin{equation}
\begin{split}
\mathbf{y}(\tau) &\triangleq \mathrm{vec} \left( \int_{T}\mathbf{x}(t,\tau){\boldsymbol \phi}^H(t)dt\right) \\
&=\displaystyle\sum_{i=1}^k\beta_i(\tau)\mathbf{a}_{\rm t}(\theta_i)\otimes\mathbf{a}_{\rm r}(\theta_i)+\tilde{\mathbf{n}}(\tau)
\end{split}
\end{equation}
where $\tilde{\mathbf{n}}(\tau)$ is the $M_{\rm t}M_{\rm r}\times1$ noise vector whose covariance is given by $\sigma_N^2\mathbf{I}_{M_{\rm t}M_{\rm r}}$, $\mathrm{vec}(\cdot)$ is the operator that stacks the columns of a matrix into one column vector, $\otimes$ denotes the Kronecker product, and $(\cdot)^H$ stands for the Hermitian transpose.

In contrast to the traditional phased-array radar, the mismatches existing in both the transmit and receive steering vectors have to be considered in MIMO radar. Then, the actual transmit and receive steering vectors can be modeled as \cite{Xiang}
\begin{equation}
\tilde{\mathbf{a}}_{\rm t} \triangleq \mathbf{a}_{\rm t} + \mathbf{e}_{\rm t},\quad \left\Vert\mathbf{e}_{\rm t}\right\Vert\leqslant\varepsilon_{\rm t}; \quad \tilde{\mathbf{a}}_{\rm r} \triangleq \mathbf{a}_{\rm r} + \mathbf{e}_{\rm r},\quad \left\Vert\mathbf{e}_{\rm r}\right\Vert\leqslant\varepsilon_{\rm r}
\end{equation}
where $\mathbf{e}_{\rm t}$ 
and $\mathbf{e}_{\rm r}$ are unknown complex vectors describing the transmit and receive steering vector mismatches, respectively, and $\left\Vert \cdot \right\Vert$ denotes the Euclidean norm of a vector.

In the mismatched case, the actual virtual steering vector can be constructed as
\begin{equation}
\begin{split}
 \tilde{\mathbf{d}} &= \tilde{\mathbf{a}}_{\rm t}\otimes\tilde{\mathbf{a}}_{\rm r} = \left(\mathbf{a}_{\rm t} + \mathbf{e}_{\rm t}\right)\otimes \left(\mathbf{a}_{\rm r} + \mathbf{e}_{\rm r}\right)\\
&=\mathbf{a}_{\rm t}\otimes\mathbf{a}_{\rm r} + \mathbf{a}_{\rm t}\otimes\mathbf{e}_{\rm r} + \mathbf{e}_{\rm t}\otimes\mathbf{a}_{\rm r} + \mathbf{e}_{\rm t}\otimes\mathbf{e}_{\rm r}.
\end{split}
\end{equation} 
The bound on the norm of the virtual steering vector mismatch can be  then found as 
\begin{equation}\label{eq: bound}
\begin{split}
\left\Vert\mathbf{e}\right\Vert &= \left\Vert\mathbf{a}_{\rm t}\otimes\mathbf{e}_{\rm r} + \mathbf{e}_{\rm t}\otimes\mathbf{a}_{\rm r} + \mathbf{e}_{\rm t}\otimes\mathbf{e}_{\rm r}\right\Vert \\
 &\leqslant \left\Vert\mathbf{a}_{\rm t}\otimes\mathbf{e}_{\rm r}\right\Vert + \left\Vert\mathbf{a}_{\rm t}\otimes\mathbf{e}_{\rm r}\right\Vert + \left\Vert\mathbf{a}_{\rm t}\otimes\mathbf{e}_{\rm r}\right\Vert \\
 &\leqslant \sqrt{M_{\rm t}}\varepsilon_{\rm r} + \sqrt{M_{\rm r}}\varepsilon_{\rm t} + \varepsilon_{\rm t}\varepsilon_{\rm r} \triangleq \varepsilon.
\end{split}
\end{equation}
It can be seen that the bound \eqref{eq: bound} which is composed of three terms is much larger than that of $\mathbf{e}_{\rm t}$ and $\mathbf{e}_{\rm r}$ considered separately. Thus, the worst-case approach which uses the loose bound \eqref{eq: bound} may be too ovely conservative. 

\section{Probability-Constrained Optimization}
Since the worst-case approach may be too overly conservative, especially in MIMO radar context, we suggest to use the probability-constrained optimization-based approach \cite{Vorobyov3}-\cite{Vorobyov6}. The key idea of this approach is to maintain the beamformer distortionless response only for operational conditions which occur with a sufficiently high probability rather than for all operational conditions corresponding to the uncertainty set. Then the joint transmit/receive robust adaptive beamforming problem for MIMO radar can be formulated as 
\begin{equation}
\begin{split}\label{eq:Probability problem}
\min_\mathbf{w}\quad &\mathbf{w}^H\hat{\mathbf{R}}\mathbf{w} \\
\textrm{s.t.}  \quad &\mathrm{Pr} \lbrace \mathbf{w}^H \left( \tilde{\mathbf{a}}_{\rm t}\otimes\tilde{\mathbf{a}}_{\rm r}\right) \geqslant 1 \rbrace \geqslant \mathnormal{p}
\end{split}
\end{equation}
where $\mathnormal{p}$ is a certain probability value, which can be selected according to the quality of service (QoS) requirements, $\mathrm{Pr} \lbrace \cdot \rbrace$ stands for the probability operator, $\mathbf{e}_{\rm t}$ 
and $\mathbf{e}_{\rm r}$ are assumed to be random, and $\hat{\mathbf{R}} \triangleq \frac{1}{L}\displaystyle\sum_{\tau=1}^L\mathbf{y(\tau)y^H(\tau)}$ is the sample covariance matrix used in practical applications. Here $L$ is the training sample size.

The general probability-constrained problem \eqref{eq:Probability problem} is nonconvex and NP-hard. However, thanks to the MIMO radar structure, the joint transmit/receive beamforming vector has the following Kronecker form $\mathbf{w} = \mathbf{u} \otimes \mathbf{v} $. Thus, we can reformulate the cost function of \eqref{eq:Probability problem} into a bi-quadratic cost function.

Using the equalities $\mathbf{a} \otimes \mathbf{b} = \mathrm{vec}\left( \mathbf{b} \mathbf{a}^T\right) $ and $\mathrm{vec}\left( \mathbf{A}\right)^H \cdot \mathrm{vec}\left( \mathbf{B}\right) = \mathrm{tr} \left( \mathbf{A}^H \mathbf{B}\right)$, we obtain that
$\mathbf{w}^H \left( \tilde{\mathbf{a}}_{\rm t} \otimes \tilde{\mathbf{a}}_{\rm r} \right) = \left( \mathbf{u}^H\tilde{\mathbf{a}}_{\rm t}\right) \left( \mathbf{v}^H \tilde{\mathbf{a}}_{\rm r}\right) $.
Then the probability constraint in \eqref{eq:Probability problem} can be reformulated as
\begin{equation}\label{eq:probability1}
\begin{split}
\mathrm{Pr} \lbrace\mathbf{w}^H  \left( \tilde{\mathbf{a}}_{\rm t} \otimes  \tilde{\mathbf{a}}_{\rm r} \right) \geqslant  1\rbrace = \mathrm{Pr} \lbrace \lvert \mathbf{u}^H\tilde{\mathbf{a}}_{\rm t} \rvert \lvert \mathbf{v}^H \tilde{\mathbf{a}}_{\rm r} \rvert \geqslant 1  \rbrace \\
\geqslant \mathrm{Pr}\lbrace \lvert \mathbf{u}^H\tilde{\mathbf{a}}_{\rm t} \rvert \geqslant 1  \cap  \lvert \mathbf{v}^H \tilde{\mathbf{a}}_{\rm r} \rvert \geqslant 1  \rbrace \geqslant \mathnormal{p}.
\end{split}
\end{equation}
Since $ \mathbf{e}_{\rm t} $ and $ \mathbf{e}_{\rm r} $ are independent identically distributed (i.i.d.), and using the fact that any functions of independent random variables are statistically independent  \cite{Papoulis}, we can split the constraint \eqref{eq:probability1} into two constraints as
\begin{equation}\label{eq:constrain}
\mathrm{Pr}\lbrace  \lvert \mathbf{u}^H\tilde{\mathbf{a}}_{\rm t}  \rvert \geqslant 1\rbrace\geqslant \eta_1, \quad
\mathrm{Pr}\lbrace  \lvert \mathbf{v}^H \tilde{\mathbf{a}}_{\rm r} \rvert \geqslant 1\rbrace\geqslant \eta_2.
\end{equation}
The probability values $\eta_1$ and $\eta_2$ are chosen separately at the transmitter and receiver so that $\eta_1 \eta_2 = \mathnormal{p}$ and $\eta_1,\eta_2\leqslant1$. Note that $\eta_1$ and $\eta_2$ are preselected based on the robustness requirements to mismatches at the transmitter and receiver. Thus, $\eta_1$ and $\eta_2$ are not the optimization variables. For example, $\eta_1=\eta_2=\sqrt{\mathnormal{p}}$ satisfies the conditions.

Let us obtain a simplified approximate form of the probability constraints \eqref{eq:constrain}. If  $\lvert \mathbf{u}^H\mathbf{a}_{\rm t}\rvert > \lvert\mathbf{u}^H\mathbf{e}_{\rm t}\rvert $ and $\lvert \mathbf{v}^H\mathbf{a}_{\rm r}\rvert > \lvert\mathbf{v}^H\mathbf{e}_{\rm r}\rvert $, i.e., if the steering vector mismatches are reasonably small, then from the triangle inequality it follows that
\begin{equation}\label{eq:triangle inequality}
\begin{split}
\lvert \mathbf{u}^H \left( \mathbf{a}_{\rm t} + \mathbf{e}_{\rm t}\right) \rvert &\geqslant \lvert \mathbf{u}^H \mathbf{a}_{\rm t}\rvert - \lvert \mathbf{u}^H \mathbf{e}_{\rm t}\rvert\\
\lvert \mathbf{v}^H \left( \mathbf{a}_{\rm r} + \mathbf{e}_{\rm r}\right) \rvert &\geqslant \lvert \mathbf{v}^H \mathbf{a}_{\rm r}\rvert - \lvert \mathbf{v}^H \mathbf{e}_{\rm r}\rvert.
\end{split}
\end{equation}
Using \eqref{eq:constrain} and \eqref{eq:triangle inequality}, the problem \eqref{eq:Probability problem} can be reformulated as
\begin{equation}\label{eq:double convex}
\begin{split}
\min_{\mathbf{u},\mathbf{v}} \quad &\left( \mathbf{u} \otimes \mathbf{v}\right)^H \hat{\mathbf{R}}\left( \mathbf{u} \otimes \mathbf{v}\right)\\
\textrm{s.t.}  \quad &\mathrm{Pr}\left\lbrace  \lvert \mathbf{u}^H\mathbf{e}_{\rm t} \rvert \leqslant \lvert \mathbf{u}^H\mathbf{a}_{\rm t} \rvert - 1 \right\rbrace\geqslant \eta_1 \\
&\mathrm{Pr}\left\lbrace  \lvert \mathbf{v}^H\mathbf{e}_{\rm r} \rvert \leqslant \lvert \mathbf{v}^H\mathbf{a}_{\rm r} \rvert - 1 \right\rbrace\geqslant \eta_2.
\end{split}
\end{equation}

The problem \eqref{eq:double convex} becomes mathematically tractable if we additionally assume a specific analytic form for the probability operator $\mathrm{Pr}\lbrace\cdot\rbrace$ and make some approximations. In the sequel, we will consider two practically important cases corresponding to two different assumptions on the probability density function (pdf) of the transmit and receive steering vector mismatches $\mathbf{e}_{\rm t}$ and $\mathbf{e}_{\rm r}$.

\subsection{Mismatches With Gaussian Distribution}
Consider the case of a complex zero-mean symmetric Gaussian
distribution for $\mathbf{e}_{\rm t}$ and $\mathbf{e}_{\rm r}$, i.e.,
\begin{equation}
\mathbf{e}_{\rm t} \sim \mathcal{CN}\left(\mathbf{0}_{M_{\rm t}},\mathbf{C}_{\rm t} \right),\quad
\mathbf{e}_{\rm r} \sim \mathcal{CN}\left(\mathbf{0}_{M_{\rm r}},\mathbf{C}_{\rm r} \right)
\end{equation}
where $\mathbf{0}_{M_{\rm t}}$ and $\mathbf{0}_{M_{\rm r}}$ denote the $M_{\rm t}\times1$ and $M_{\rm r}\times1$ vector of zeros, respectively, $\mathbf{C}_{\rm t}$ and $\mathbf{C}_{\rm r}$ capture the second-order statistics of the uncertainties in the transmit and receive steering vectors, respectively. .

We consider the receive steering vector mismatch and note that the same considerations apply to the transmit steering vector mismatch. It is easy to show that the random variable $\mathbf{v}^H\mathbf{e}_{\rm r}$ has the complex Gaussian distribution, that is, $\mathbf{v}^H\mathbf{e}_{\rm r} \sim \mathcal{CN}\left(\mathbf{0}_{M_{\rm r}},\mathbf{v}^H\mathbf{C}_{\rm r}\mathbf{v} \right)$, and its real and imaginary parts are real i.i.d. Gaussian. 
 
Let us use the fact that if $ \mathnormal{x} $ and $ \mathnormal{y} $ are two real i.i.d. zero mean Gaussian random variables with the variance $\sigma^2$, then $ \mathnormal{z} = \sqrt{\mathnormal{x}^2 + \mathnormal{y}^2} $ is Rayleigh-distributed with the cumulative density function (cdf) given as $\mathnormal{F}(\mathnormal{z}) = 1 - \mathnormal{e}^{-\mathnormal{z}^2/2\sigma^2}$.
Using this fact, the probability constraint \eqref{eq:constrain} at the receiver can be written as 
\begin{equation}\label{eq: probability constrain}
\begin{split}
\mathrm{Pr} &\left\lbrace  \lvert \mathbf{v}^H\mathbf{e}_{\rm r} \rvert \leqslant \lvert \mathbf{v}^H\mathbf{a}_{\rm r} \rvert - 1 \right\rbrace \\
&= 1 - \mathnormal{exp}\left( - \frac{(\lvert \mathbf{v}^H\mathbf{a}_{\rm r} \rvert - 1)^2}{\mathbf{v}^H\mathbf{C}_{\rm r}\mathbf{v}}\right) \geqslant \eta_2.
\end{split}
\end{equation} 
The inequality in the second line of \eqref{eq: probability constrain} can be equivalently rewritten as
\begin{equation}
\lvert \mathbf{v}^H\mathbf{a}_{\rm r} \rvert - \sqrt{\mathnormal{ln}\left( \frac{1}{1- \eta_2}\right)}\left\Vert \mathbf{C}_{\rm r}^{\frac{1}{2}}\mathbf{v}\right\Vert \geqslant 1.
\end{equation} 
 
Observing that the cost function in \eqref{eq:double convex} is unchanged when $\mathbf{u}$ and $\mathbf{v}$ undergo an arbitrary phase rotation, the problem \eqref{eq:double convex} can be further rewritten as
\begin{equation}\label{eq:probability problem 2}
\begin{split}
\min_{\mathbf{u},\mathbf{v}} \quad &\left( \mathbf{u} \otimes \mathbf{v}\right)^H \hat{\mathbf{R}}\left( \mathbf{u} \otimes \mathbf{v}\right)\\
\textrm{s.t.}  \quad  &\mathbf{u}^H\mathbf{a}_{\rm t} - \gamma_{11}\left\Vert \mathbf{C}_{\rm t}^{\frac{1}{2}}\mathbf{u}\right\Vert \geqslant 1,\quad \mathcal{I}m(\mathbf{u}^H\mathbf{a}_{\rm t})=0\\
&\mathbf{v}^H\mathbf{a}_{\rm r} - \gamma_{12}\left\Vert \mathbf{C}_{\rm r}^{\frac{1}{2}}\mathbf{v}\right\Vert \geqslant 1, \quad \mathcal{I}m(\mathbf{v}^H\mathbf{a}_{\rm r})=0 \\
\end{split}
\end{equation} 
where $\gamma_{11} \triangleq \sqrt{\mathnormal{ln}( (1- \eta_1)^{-1}) },$ and $\gamma_{12} \triangleq \sqrt{\mathnormal{ln}((1- \eta_2)^{-1}) }$.

The problem \eqref{eq:probability problem 2} has separate constraints for $\mathbf{u}$ and $\mathbf{v}$, but coupled objective. For such type of problems of minimizing a continuous function of several blocks of variables, the block coordinate descent (BCD) methods are widely used \cite{Tseng}-\cite{Zou}. At each iteration of BCD, a single block of variables is optimized, while the remaining variables are held fixed. Thus, we develop here a BCD type method for addressing the problem \eqref{eq:probability problem 2}. It can be seen that if one of the vectors $\mathbf{u}$ or $\mathbf{v}$ is
fixed, the cost function of the problem \eqref{eq:probability problem 2} can be transformed into a quadratic function with respect to the other vector. Hence, it can be solved based on the second-order cone programming (SOCP) using, for example, the CVX toolbox \cite{CVX}. 

Let us fix $\mathbf{u}$ and choose $ \mathbf{u}_{\mathnormal{k}} = \mathbf{u}_{\mathnormal{k-1}}^{\mathnormal{opt}}$. It is obvious that there is a scaling determinacy between $\mathbf{u}$ and $\mathbf{v}$. Then, \eqref{eq:probability problem 2} boils down to 
\begin{equation}\label{eq:probability problem 3}
\begin{split}
\min_{\mathbf{v}} \quad &\left( \mathbf{u}_{\mathnormal{k}} \otimes \mathbf{v}\right)^H \hat{\mathbf{R}}\left( \mathbf{u}_{\mathnormal{k}} \otimes \mathbf{v}\right)\\
\textrm{s.t.}  \quad  & \left( \mathbf{u}_{\mathnormal{k}}^H\mathbf{a}_{\rm t} - \gamma_{11}\left\Vert \mathbf{C}_{\rm t}^{\frac{1}{2}}\mathbf{u}_{\mathnormal{k}}\right\Vert\right) \left( \mathbf{v}^H\mathbf{a}_{\rm r} - \gamma_{12}\left\Vert \mathbf{C}_{\rm r}^{\frac{1}{2}}\mathbf{v}\right\Vert\right) \geqslant 1\\
&\mathcal{I}m(\mathbf{v}^H\mathbf{a}_{\rm r} ) = 0
\end{split}
\end{equation}
By solving \eqref{eq:probability problem 3}, we obtain the optimal solution $\mathbf{v}_{k-1}^{opt}$. Then we fix $\mathbf{v}$ and choose $ \mathbf{v}_{\mathnormal{k}} = \mathbf{v}_{\mathnormal{k-1}}^{\mathnormal{opt}}$. The problem \eqref{eq:probability problem 2} boils down respectively to 
\begin{equation}\label{eq:probability problem 4}
\begin{split}
\min_{\mathbf{u}} \quad &\left( \mathbf{u} \otimes \mathbf{v}_{\mathnormal{k}}\right)^H \hat{\mathbf{R}}\left( \mathbf{u} \otimes \mathbf{v}_{\mathnormal{k}}\right)\\
\textrm{s.t.}  \quad  &\left( \mathbf{u}^H\mathbf{a}_{\rm t} - \gamma_{11}\left\Vert \mathbf{C}_{\rm t}^{\frac{1}{2}}\mathbf{u}\right\Vert\right) \left( \mathbf{v}_{\mathnormal{k}}^H\mathbf{a}_{\rm r} - \gamma_{12}\left\Vert \mathbf{C}_{\rm r}^{\frac{1}{2}}\mathbf{v}_{\mathnormal{k}}\right\Vert\right) \geqslant 1\\
&\mathcal{I}m(\mathbf{u}^H\mathbf{a}_{\rm t}) = 0.
\end{split}
\end{equation}
The overall algorithm for addressing \eqref{eq:probability problem 2} is then given as in Algorithm 1. The convergence of a large class of BCD methods is investigated in \cite{Razaviyayn}.

\subsection{Mismatch With Arbitrary Distribution}
Consider now the case when the transmit and receive steering vector mismatches are arbitrary distributed and only the first and second-order statistics are known.
\begin{algorithm}[t]
\caption{ :~~Interative SOCP Method}    
\begin{algorithmic}[1]
\STATE \textbf{Given:} initial value $ \mathbf{u}_0 = \frac{\mathbf{a}_{\rm t}}{\left\Vert\mathbf{a}_{\rm t}\right\Vert} $, $\mathnormal{k}=0$, tolerance $\delta$ \\
\textbf{Repeat:} $\mathnormal{k} = \mathnormal{k}+1$
\STATE Solve \eqref{eq:probability problem 3}, obtain $\mathbf{v}_{\mathnormal{k-1}}^{\mathnormal{opt}}, \,\mathbf{v}_{\mathnormal{k}}: = \mathbf{v}_{\mathnormal{k-1}}^{\mathnormal{opt}}$
\STATE Solve \eqref{eq:probability problem 4}, obtain $\mathbf{u}_{\mathnormal{k-1}}^{\mathnormal{opt}}, \,\mathbf{u}_{\mathnormal{k}}: = \mathbf{u}_{\mathnormal{k-1}}^{\mathnormal{opt}}$ \\
\textbf{Until} 
\STATE $ \frac{\left\Vert \mathbf{u}_{\mathnormal{k-1}}^{\mathnormal{opt}} - \mathbf{u}_{\mathnormal{k}-1}\right\Vert}{\left\Vert \mathbf{u}_{\mathnormal{k-1}}^{\mathnormal{opt}}\right\Vert} < \delta $ and $ \frac{\left\Vert \mathbf{v}_{\mathnormal{k-1}}^{\mathnormal{opt}} - \mathbf{v}_{\mathnormal{k}-1}\right\Vert}{\left\Vert \mathbf{v}_{\mathnormal{k-1}}^{\mathnormal{opt}}\right\Vert} < \delta $\\
\textbf{Return}
\STATE \textbf{Output:} $ \mathbf{u}^{\mathnormal{opt}}$ and  $\mathbf{v}^{\mathnormal{opt}}$.
\end{algorithmic}
\end{algorithm}

Using the Chebyshev inequality which states that for any zero-mean random variable $\tau$ with variance $\sigma_{\tau}^{2}$ and positive real number $\alpha$, we obtain $\mathrm{Pr}\lbrace\lvert \tau\rvert \geqslant\alpha\rbrace \leqslant \frac{\sigma_{\tau}^{2}}{\alpha^2}$. Then the probability constraint \eqref{eq:constrain} at the receiver can be expressed as
\begin{equation}\label{eq:constrain 1}
\mathrm{Pr} \left\lbrace  \lvert \mathbf{v}^H\mathbf{e}_{\rm r} \rvert \leqslant \lvert \mathbf{v}^H\mathbf{a}_{\rm r} \rvert - 1 \right\rbrace \\
= 1 - \frac{\mathbf{v}^H\mathbf{C}_{\rm r}\mathbf{v}} {(\lvert \mathbf{v}^H\mathbf{a}_{\rm r} \rvert - 1)^2}\geqslant \eta_2.
\end{equation}
Further, \eqref{eq:constrain 1} can be rewritten at both the transmit and receive sides as 
\begin{equation}
\lvert \mathbf{u}^H\mathbf{a}_{\rm t} \rvert - \gamma_{21}\left\Vert \mathbf{C}_{\rm t}^{\frac{1}{2}}\mathbf{u}\right\Vert \geqslant 1,\quad\lvert \mathbf{v}^H\mathbf{a}_{\rm r} \rvert - \gamma_{22}\left\Vert \mathbf{C}_{\rm r}^{\frac{1}{2}}\mathbf{v}\right\Vert \geqslant 1
\end{equation}
where $\gamma_{21} \triangleq \left( \sqrt{1- \eta_1}\right) ^{-1},$ and $\gamma_{22} \triangleq \left( \sqrt{1- \eta_2}\right) ^{-1}$. We can see that the difference between the cases of the mismatch with arbitrary and Gaussian distribution is in the coefficients $\gamma_{11},\gamma_{12}$ and $\gamma_{21},\gamma_{22}$ only. Thus, the problem in the case of the mismatch with arbitrary distribution can be addressed in the same way as before.

\section{Tight Lower Bound for The Probability Constraint Using Duality Theory}
For simplicity, let us eliminate the subscripts in our notations and instead use  simply the notations $\mathbf{a}$, $\mathbf{e}$, and $\mathbf{C}$. The tight lower bound problem is
\begin{equation}\label{eq:Chebyshev inequality}
\min_{\mathnormal{f}(\mathbf{e})}\quad \mathrm{Pr} \lbrace \lvert \mathbf{v}^H(\mathbf{a}+\mathbf{e}) \rvert \geqslant 1 \rbrace
\end{equation}
where  $\mathnormal{f}(\mathbf{e})$ is the pdf of $\mathbf{e}$. It can be reformulated as  \cite{He}
\begin{equation}\label{eq:lower level problem}
\begin{split}
\min_{\mathnormal{f}(\mathbf{e})}\quad &\mathrm{Pr} \lbrace \mathnormal{g}(\mathbf{e}) \geqslant 1 \rbrace \\
\textrm{s.t.}\quad &\mathbb{E}\lbrace\mathbf{e}\mathbf{e}^H\rbrace = \mathbf{C} \\
&\int_{\mathbf{e}}{\mathnormal{f}(\mathbf{e})\mathnormal{d}\mathbf{e}} = 1,\quad\mathbb{E}\lbrace\mathbf{e}\rbrace = 0
\end{split}
\end{equation}
where $\mathnormal{g}(\mathbf{e}) \triangleq \lvert \mathbf{v}^H(\mathbf{a}+\mathbf{e}) \rvert^2$. To find the dual problem for \eqref{eq:lower level problem}, we first introduce the Lagrangian function as 
\begin{equation*}
\begin{split}
&\mathbb{L}(\mathnormal{f}(\mathbf{e}),\mu,{\boldsymbol \eta},\mathbf{T})= \mathrm{Pr} \lbrace \mathnormal{g}(\mathbf{e}) \geqslant 1 \rbrace + \mu(1 - \int_{\mathbf{e}}{\mathnormal{f}(\mathbf{e})\mathnormal{d}\mathbf{e}}) \\
&\quad + {\boldsymbol \eta}^H(\mathbf{0} - \mathbb{E}\lbrace\mathbf{e}\rbrace) + \mathrm{tr}(\mathbf{T}^H(\mathbf{C} - \mathbb{E}\lbrace\mathbf{e}\mathbf{e}^H\rbrace))\\
&= \mu + \mathrm{tr}(\mathbf{T}^H\mathbf{C}) \\
\end{split} 
\end{equation*}
\begin{equation}\label{eq:lower level problem1}
\begin{split}
&\quad +\int_{\mathnormal{g}(\mathbf{e})\geqslant 1} [ 1 - (\mu + {\boldsymbol \eta}^H\mathbf{e} + \mathrm{tr}(\mathbf{T}^H\mathbf{e}\mathbf{e}^H)) ] {\mathnormal{f}(\mathbf{e})\mathnormal{d}\mathbf{e}}\\&\quad + \int_{\mathnormal{g}(\mathbf{e})\leqslant 1} [ 0 - (\mu + {\boldsymbol \eta}^H\mathbf{e} + \mathrm{tr}(\mathbf{T}^H\mathbf{e}\mathbf{e}^H)) ] {\mathnormal{f}(\mathbf{e})\mathnormal{d}\mathbf{e}}
\end{split} 
\end{equation}
where $\mu$, ${\boldsymbol \eta}$, and $\mathbf{T}$ are the Lagrange multipliers and $\mathbf{T} = \mathbf{T}^H$. With the implicit pdf constraint $\mathnormal{f}(\mathbf{e})\geqslant 0$, the Lagrange dual function of the problem \eqref{eq:lower level problem} is given as
\begin{equation}
g(\mu,{\boldsymbol \eta},\mathbf{T}) = \min_{\mathnormal{f}(\mathbf{e})\geqslant 0}\mathbb{L}(\mathnormal{f}(\mathbf{e}),\mu,{\boldsymbol \eta},\mathbf{T}).
\end{equation}

Note that the minimum of the first integral in \eqref{eq:lower level problem1} with the nonnegative pdf $\mathnormal{f}(\mathbf{e})$ constrain is zero if $\mu + {\boldsymbol \eta}^H\mathbf{e} + \mathrm{tr}(\mathbf{T}^H\mathbf{e}\mathbf{e}^H)\leqslant 1$, otherwise the minimum is unbounded below. Similarly, the minimum of the second integral with the nonnegative pdf $\mathnormal{f}(\mathbf{e})$ constrain is zero if $\mu + {\boldsymbol \eta}^H\mathbf{e} + \mathrm{tr}(\mathbf{T}^H\mathbf{e}\mathbf{e}^H)\leqslant 0$, otherwise the minimum is unbounded below. It is easy to see that the condition $\mu + {\boldsymbol \eta}^H\mathbf{e} + \mathrm{tr}(\mathbf{T}^H\mathbf{e}\mathbf{e}^H)\leqslant 0$ must hold for both $\mathnormal{g}(\mathbf{e})\geqslant 1$ and $\mathnormal{g}(\mathbf{e})\leqslant 1$, i.e., $\forall \mathnormal{g}(\mathbf{e})$, and therefore, $\forall\mathbf{e}$, while the condition $\mu + {\boldsymbol \eta}^H\mathbf{e} + \mathrm{tr}(\mathbf{T}^H\mathbf{e}\mathbf{e}^H)\leqslant 0$ must hold for $\forall \mathnormal{g}(\mathbf{e})\leqslant 1$. Then the dual problem to \eqref{eq:lower level problem} can be formulated as 
\begin{equation}\label{eq: problem 27}
\begin{split}
\max_{\mu,{\boldsymbol \eta},\mathbf{T}} \quad &\mu + \mathrm{tr}(\mathbf{T}^H\mathbf{C}) \\
\textrm{s.t.} \quad &\mu + {\boldsymbol \eta}^H\mathbf{e} + \mathrm{tr}(\mathbf{T}^H\mathbf{e}\mathbf{e}^H)\leqslant 1, \quad \forall \mathbf{e}\\
& \mu + {\boldsymbol \eta}^H\mathbf{e} + \mathrm{tr}(\mathbf{T}^H\mathbf{e}\mathbf{e}^H)\leqslant 0, \quad \forall \mathnormal{g}(\mathbf{e})\leqslant 1 \\
& \mathbf{T} = \mathbf{T}^H.
\end{split}
\end{equation}

Let us define $$\tilde{\mathbf{e}}\triangleq\begin{bmatrix}\mathbf{e} \\ 1 \end{bmatrix},\widetilde{\mathbf{C}} \triangleq \begin{bmatrix}
 \mathbf{C}&\mathbf{0}\\ \mathbf{0}^H & 1
 \end{bmatrix}, \mathbf{Z} \triangleq \begin{bmatrix}
 \mathbf{T}^H & \frac{1}{2}{\boldsymbol \eta}^H \\ \frac{1}{2}{\boldsymbol \eta}^H &\mu
\end{bmatrix}, $$
$$
\mathbf{A} \triangleq \begin{bmatrix}
\mathbf{v}\mathbf{v}^H &\mathbf{v}\mathbf{v}^H\mathbf{a} \\ \mathbf{a}^H\mathbf{v}\mathbf{v}^H &\mathbf{a}^H\mathbf{v}\mathbf{v}^H\mathbf{a}-1
\end{bmatrix}.  $$
Thus, the second constrain in \eqref{eq: problem 27} can be rewritten as $\tilde{\mathbf{e}}^H\mathbf{Z}\tilde{\mathbf{e}}\leqslant 0, \forall \tilde{\mathbf{e}}^H\mathbf{A}\tilde{\mathbf{e}} \leqslant 0$. Further, according to \textit{S-Lemma} it becomes $\mathbf{Z} - \lambda \mathbf{A} \preccurlyeq \mathbf{0}, \forall \lambda \geqslant 0 $.  Then the problem \eqref{eq: problem 27} in compact form can be expressed as
\begin{equation}
\begin{split}
\max_{\mathbf{Z}} \quad &\mathrm{tr}(\mathbf{Z}\widetilde{\mathbf{C}}) \\
\textrm{s.t.} \quad &\mathbf{Z}-\begin{bmatrix}
\mathbf{0} &\mathbf{0} \\ \mathbf{0}^H &1
\end{bmatrix}\preccurlyeq\mathbf{0}, \quad \mathbf{Z} = \mathbf{Z}^H\\
& \mathbf{Z} - \lambda \mathbf{A} \preccurlyeq \mathbf{0}, \quad \forall \lambda \geqslant 0.
\end{split}
\end{equation}
Since the Lagrange dual problem yields the lower bound of the primal problem, we can finally state that 
\begin{equation}
\min_{\mathnormal{f}(\mathbf{e})}\:\mathrm{Pr} \lbrace \mathnormal{g}(\mathbf{e}) \geqslant 1 \rbrace  =
\max_{\mathbf{Z}} \;\mathrm{tr}(\mathbf{Z}\widetilde{\mathbf{C}}).
\end{equation}

\section{Simulation Results}
We assume a uniform linear array (ULA) of $10$ antenna elements that is used for both transmitting and receiving, i.e., $M_{\rm t} = 10$ and $M_{\rm r} = 10$.  Antenna elements are spaced half a wavelength apart from each other. The plane-wave target  impinges on the array from $\theta_{\rm s}=3^\circ$ and two interfering sources arrive from directions $\theta_{\rm i1}=30^\circ$ and $\theta_{\rm i2}=50^\circ$, respectively. The interference-to-noise ratio (INR) is assumed to be $20$ $dB$ for both interferences, and $100$ Monte-Carlo runs are used to obtain each point in our simulations curves.

Consider the scenario with Ricean propagation medium where the mismatch vectors $\mathbf{e}_{\rm t}$ and $\mathbf{e}_{\rm r}$ are modelled as
\[
\mathbf{e} = \frac{\sigma}{\sqrt{MN}}\sum_{n=1}^{N}e^{j\psi_n}\mathbf{a}(\theta_0+\theta_n)
\]
where $M$ stands for $M_{\rm t}$ or $M_{\rm r}$, $\sigma$ is the power of scattered nonline-of-sight (NLOS) components, $N$ is the number of NLOS components, $\psi_n$ is the phase shift parameter of the $n$th NLOS component, and $\theta_n$ is angular shift. The parameters $\theta_n$ and $\psi_n,n=1,\ldots,N$ are independently and uniformly drawn in each simulation run from $[-2.5^\circ, 2.5^\circ]$ and $[0, 2\pi)$, respectively. The values $N=10$ and $\sigma^2=0.3 M$ are taken.
 
For comparison, $\varepsilon = 9$  is chosen for the worst-case robust beamformer \cite{Xiang}. For the LSMI beamformer, the fixed diagonal loading parameter $\gamma = 10$ is chosen. For the proposed probability-constrained optimization based joint transmit/receive robust adaptive beamformer, $\mathnormal{p} = 0.9$ while $\eta_1=0.93$ and $\eta_2=\mathnormal{p}/\eta_1=0.9677$ are taken.
\begin{figure}[t]
\centerline{\includegraphics[width=15.5cm]{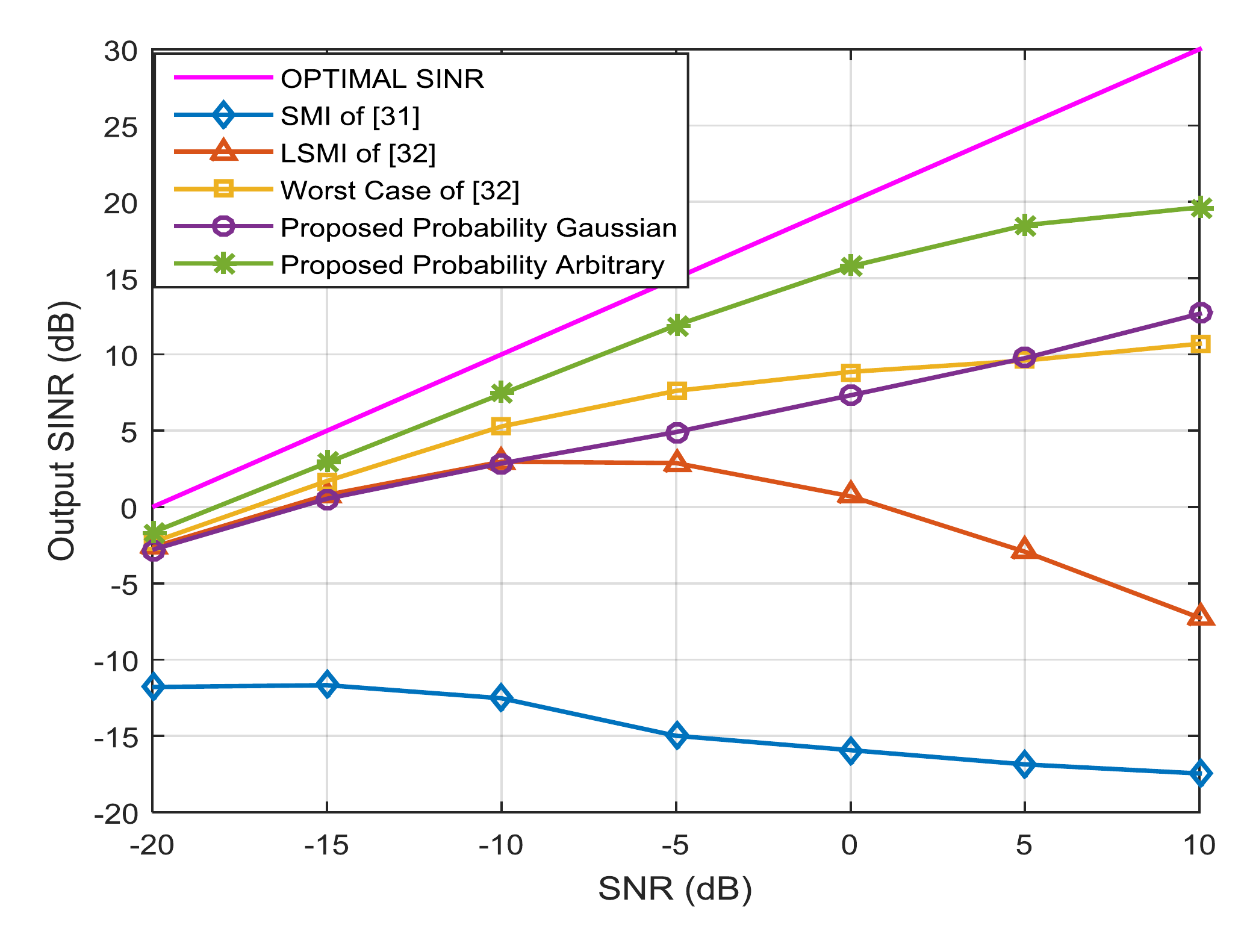}}
\caption{Output SINR versus SNR for K=100 and INR=20dB.} \label{fi:Fig1}
\end{figure}

The output SINRs versus SNR are shown in Fig.~\ref{fi:Fig1}. We can see that the proposed method based on the probability-constrained optimization for the case of arbitrary distributed mismatches has the best performance among all the techniques tested. These improvement is especially remarkable at high SNRs. However, the performance of the probability-constrained optimization based method with Gaussian mismatch distribution is worse than that for the worst-case robust beamformer. It can be explained by the fact that the actual mismatch corresponding to the considered Ricean scenario is not Gaussian.

\section{Conclusion}
A joint transmit/receive robust adaptive beamforming method for MIMO radar has been developed based on the probability-constrained optimization. Specifically, we have considered the mismatches in the desired signal steering vectors at both transmit and receive arrays to be random. The original probability-constrained optimization problem has been converted into a biconvex problem and addressed by using BCD approach. The proposed technique offers a better performance than several state-of-the-art counter parts.

\end{document}